\documentclass[submission,copyright,creativecommons]{eptcs}
\providecommand{\event}{QPL 2014} 
\usepackage{breakurl}             
\title{Quantum Set Theory Extending the Standard Probabilistic Interpretation of Quantum Theory (Extended Abstract)}
\author{Masanao Ozawa\thanks{Supported by JSPS KAKENHI, No. 26247016,
and the John Templeton Foundation, ID \#35771.}
\institute{Graduate School of Information Science, Nagoya University\\ 
Nagoya, Japan}
\email{ozawa@is.nagoya-u.ac.jp}
}
\def\titlerunning{Quantum set theory}
\def\authorrunning{M. Ozawa}
\usepackage{graphicx}
\usepackage{amsthm}
\usepackage{mathptmx}
\usepackage{latexsym}
\usepackage{amsmath,amssymb,graphicx,color,bm}
\usepackage{enumerate}

\newcommand{\af}{\,\et\,}
\newcommand{\al}{\alpha}
\newcommand{\be}{\beta}
\newcommand{\benum}[1]{\begin{enumerate}[#1]\itemsep=0in}
\newcommand{\beq}{\begin{equation}}
\newcommand{\beqa}{\begin{eqnarray}}
\newcommand{\beqas}{\begin{eqnarray*}}
\newcommand{\beql}[1]{\begin{equation}\label{eq:#1}}
\newcommand{\bitem}{\begin{itemize}\itemsep=0in}
\newcommand{\bM}{{\bf M}}
\newcommand{\bP}{{\bf P}}
\newcommand{\bra}[1]{\left\langle#1\right|}
\newcommand{\bracket}[1]{\left\langle#1\right\rangle}

\newcommand{\bS}{{\bf S}}
\newcommand{\bx}{{\bf x}}
\newcommand{\cA}{{\cal A}}
\newcommand{\cB}{{\cal B}}
\newcommand{\cC}{{\cal C}}
\newcommand{\cD}{{\cal D}}

\newcommand{\cH}{{\cal H}}

\newcommand{\cK}{{\cal K}}

\newcommand{\cL}{{\cal L}}

\newcommand{\cM}{{\cal M}}
\newcommand{\cm}{\underline{\Or}}

\newcommand{\cO}{{\cal O}}
\newcommand{\com}{{\rm com}}
\newcommand{\commutes}{\ {}^{|}\!\!\!{}_{\circ}\ }
\newcommand{\cP}{{\cal P}}
\newcommand{\cQ}{{\cal Q}}

\newcommand{\cS}{{\cal S}}

\newcommand{\cuniv}{{\rm com}}

\newcommand{\da}{\dagger}
\newcommand{\de}{\delta}
\newcommand{\De}{\Delta}
\newcommand{\dom}{\cD}
\newcommand{\eenum}{\end{enumerate}}
\newcommand{\eeq}{\end{equation}}
\newcommand{\eeqa}{\end{eqnarray}}
\newcommand{\eeqas}{\end{eqnarray*}}
\newcommand{\eitem}{\end{itemize}}

\newcommand{\Eq}[1]{Eq.~(\ref{eq:#1})}
\newcommand{\et}{\eta}

\newcommand{\Ga}{\Gamma}

\newcommand{\hu}{\hat{u}}

\newcommand{\Iff}{\leftrightarrow}
\newcommand{\Inf}{\bigwedge}

\newcommand{\ket}[1]{\left|#1\right\rangle}
\newcommand{\ketbra}[1]{\ket{#1}\!\!\bra{#1}}
\newcommand{\la}{\lambda}

\newcommand{\M}{\bM}
\newcommand{\mb}{\mbox}

\newcommand{\Not}{\neg}
\newcommand{\om}{\omega}

\newcommand{\Or}{\vee}
\newcommand{\p}{{}^{\perp}}

\newcommand{\ph}{\phi}

\newcommand{\ps}{\psi}
\newcommand{\Q}{{\bf Q}}
\newcommand{\R}{{\bf R}}

\newcommand{\rank}{\mbox{\rm rank}}
\newcommand{\RB}{\R^{(\B)}}
\newcommand{\rh}{\rho}
\newcommand{\RQ}{\R^{(\cQ)}}
\newcommand{\si}{\sigma}

\newcommand{\Sp}{{\rm Sp}}
\newcommand{\Sup}{\bigvee}

\newcommand{\tA}{\tilde{A}}

\newcommand{\Then}{\Rightarrow}
\newcommand{\Tr}{\mbox{\rm Tr}}
\newcommand{\tX}{\tilde{X}}
\newcommand{\tY}{\tilde{Y}}

\newcommand{\V}{V}
\newcommand{\val}[1]{[\![#1]\!]}
\newcommand{\valo}[1]{[\![#1]\!]_{o}}
\newcommand{\VB}{\V^{(\B)}}

\newcommand{\VL}{\V^{(\cQ)}}

\newcommand{\VQ}{\V^{(\cQ)}}

\newcommand{\vval}[1]{[\![#1]\!]_{\cQ}}

\renewcommand{\And}{\wedge}
\renewcommand{\inf}{\bigwedge}
\renewcommand{\RB}{\R^{(\cB)}}

\renewcommand{\sup}{\bigvee}
\renewcommand{\Then}{\rightarrow}
\renewcommand{\VB}{\V^{(\cB)}}
\renewcommand{\Then}{\rightarrow}
\renewcommand{\VB}{\V^{(\cB)}}
\newtheorem{Theorem}{Theorem}

\begin{document}
\maketitle

\begin{abstract}
The notion of equality between two observables will play many important roles in foundations 
of quantum theory.
However, the standard probabilistic interpretation based on the conventional Born formula
does not give the probability of equality relation for a pair of arbitrary observables, 
since the Born formula gives the probability distribution only for a commuting family of observables.
In this paper,  quantum set theory developed by Takeuti and the present author
is used to systematically extend the probabilistic interpretation of quantum theory 
to define the probability of equality relation for a pair of arbitrary observables.
Applications of this new interpretation to measurement theory are discussed briefly.
\end{abstract}

\section{Introduction}
\label{se:1}
Set theory provides foundations of mathematics.
All the mathematical notions like numbers, functions, relations, and structures
are defined in the axiomatic set theory, 
ZFC  (Zermelo-Fraenkel set theory with the axiom of choice), and 
all the mathematical theorems are required to be provable in ZFC \cite{TZ71}.
Quantum set theory instituted by Takeuti \cite{Ta81} and developed by the
present author \cite{07TPQ} naturally extends the logical basis of set theory 
from classical logic to quantum logic \cite{BvN36}.
Accordingly, quantum set theory extends quantum logical approach 
to quantum foundations from propositional logic to predicate logic and set theory. 
Hence, we can expect that quantum set theory will provide much more systematic 
interpretation of quantum theory than the conventional quantum logic approach
\cite{Gib87}.

The notion of equality between quantum observables will play many important roles 
in foundations of quantum theory, in particular, in the theory of measurement and 
disturbance \cite{04URN,06QPC}.
However, the standard probabilistic interpretation based on the conventional Born formula
does not give the probability of equality relation for a pair of arbitrary observables, 
since the Born formula gives the probability distribution only for a commuting family 
of observables \cite{vN55}.

In this paper,  quantum set theory is used to systematically extend the probabilistic 
interpretation of quantum theory to define the probability of equality relation 
for a pair of arbitrary observables, based on the fact that real numbers constructed 
in quantum set theory exactly corresponds to quantum observables \cite{Ta81,07TPQ}.  
It is shown that all the observational propositions on a quantum system correspond
to statements in quantum set theory with the same projection-valued truth value assignments
and the same probability assignments in any state.
In particular, the equality relation for real numbers in quantum set theory naturally provides 
the equality relation for quantum mechanical observables.
It has been broadly accepted that we cannot speak of the values of quantum observables 
without assuming a hidden variable theory, which severely constrained by 
Kochen-Specker type no-go theorems \cite{KS67,Red87}. 
However, quantum set theory enables us to do so without assuming
hidden variables but alternatively with the consistent use of quantum logic. 
Applications of this new interpretation to measurement theory are discussed briefly.

Section 2 provides preliminaries on commutators in complete orthomodular lattices, 
which play a fundamental role in quantum set theory.  
Section 3 introduces quantum logic on Hilbert spaces and section 4 introduces quantum
set theory and the transfer principle from theorems in ZFC to valid statements in quantum
set theory established in Ref.~\cite{07TPQ}.  Section 5 introduces the Takeuti correspondence
between reals in quantum set theory to observables in quantum theory found by 
Takeuti~\cite{Ta81}.  Section 6 formulates the standard probabilistic interpretation of quantum
theory and also shows that observational propositions for a quantum system can be embedded 
in statements in quantum set theory with the same projection-valued truth value assignment.  
Section 7 extends the standard interpretation by introducing state-dependent
joint determinateness relation.  Section 8 extends the standard interpretation by introducing 
state-dependent equality for arbitrary two observables.  Section 9 and 10 provide applications
to quantum measurement theory.

\section{Complete orthomodular lattices and commutators}
\label{se:QL}

A {\em complete orthomodular lattice}  is a complete
lattice $\cQ$ with an {\em orthocomplementation},
a unary operation $\perp$ on $\cQ$ satisfying
(C1)  if $P \le Q$ then $Q^{\perp}\le P^{\perp}$,
(C2) $P^{\perp\perp}=P$,
(C3) $P\Or P^{\perp}=1$ and $P\And P^{\perp}=0$,
where $0=\Inf\cQ$ and $1=\Sup\cQ$,
that  satisfies the {\em orthomodular law}
(OM) if $P\le Q$ then $P\Or(P^{\perp}\And Q)=Q$.
In this paper, any complete orthomodular lattice is called a {\em logic}.
A non-empty subset of a logic $\cQ$ is called a {\em subalgebra} iff
it is closed under $\And $, $\Or$, and $\perp$.
A subalgebra $\cA$ of $\cQ$ is said to be {\em complete} iff it has
the supremum and the infimum in $\cQ$ of an arbitrary subset of $\cA$.
For any subset $\cA$ of $\cQ$, 
the subalgebra generated by $\cA$ is denoted by
$\Ga_0\cA$.
We refer the reader to Kalmbach \cite{Kal83} for a standard text on
orthomodular lattices.

We say that $P$ and $Q$ in a logic $\cQ$ {\em commute}, in  symbols
$P\commutes Q$, iff  $P=(P\And Q)\Or(P\And
Q^{\perp})$.  A logic $\cQ$ is a Boolean
algebra if and only if $P\commutes Q$  for all $P,Q\in\cQ$ \cite[pp.~24--25]{Kal83}.
For any subset $\cA\subseteq\cQ$,
we denote by $\cA^{!}$ the {\em commutant} 
of $\cA$ in $\cQ$ \cite[p.~23]{Kal83}, i.e., 
\[
\cA^{!}=
\{P\in\cQ\mid P\commutes Q \mbox{ for all }
Q\in\cA\}.
\]
Then, $\cA^{!}$ is a complete subalgebra of $\cQ$.
A {\em sublogic} of $\cQ$ is a subset $\cA$ of
$\cQ$ satisfying $\cA=\cA^{!!}$. 
For any subset $\cA\subseteq\cQ$, the smallest 
logic including $\cA$ is 
$\cA^{!!}$ called the  {\em sublogic generated by
$\cA$}.
Then, it is easy to see that a subset 
 $\cA$ is a Boolean sublogic, or equivalently 
 a distributive sublogic, if and only if 
$\cA=\cA^{!!}\subseteq\cA^{!}$.

Let $\cQ$ be a logic.
Marsden \cite{Mar70} introduced the {\em commutator} $\com(P,Q)$ 
of two elements $P$ and $Q$ of $\cQ$ by 
\beqas
\com(P,Q)=(P\And Q)\Or(P\And Q\p)\Or(P\p\And Q)\Or(P\p\And Q\p).
\eeqas
Generalizing this notion to an arbitrary subset $\cA$ of $\cQ$, Takeuti \cite{Ta81} defined
the commutator $\com(\cA)$ of $\cA$ by
\beqas
\com(\cA)=\Sup\{E\in\cA^{!} \mid P_{1}\And E\commutes P_{2}\And E
\mb{ for all }P_{1},P_{2}\in\cA\}.
\eeqas
Subsequently, Chevalier \cite{Che89} proved the relation
\beqas
\com(\cA)=\Inf\{\com(P,Q)\mid P,Q\in\Ga_0(\cA)\},
\eeqas
which concludes $\com(\cA)\in\cA^{!!}\cap\cA^{!}$.
For any $P,Q\in\cQ$, the {\em interval} $[P,Q]$ is the set of
all $X\in\cQ$ such that $P\le X\le Q$.
For any $\cA\subseteq \cQ$ and $P,Q\in\cA$,
we write $[P,Q]_{\cA}=[P,Q]\cap\cA$.
The following theorem clarifies the significance of commutators.

\begin{Theorem}\label{th:absoluteness_commutator}
Let $\cA$ be a subset of a logic $\cQ$.
Then,  $\cA^{!!}$ is isomorphic to the direct product of the complete Boolean algebra
$[0,\com(\cA)]_{\cA^{!!}}$ and the complete orthomodular lattice
$[0,\com(\cA)^\perp]_{\cA^{!!}}$ without non-trivial Boolean factor.
\end{Theorem}

\section{Quantum logic on Hilbert spaces}
\label{se:QL}

Let $\cH$ be a Hilbert space.
For any subset $S\subseteq\cH$,
we denote by $S^{\perp}$ the orthogonal complement
of $S$.
Then, $S^{\perp\perp}$ is the closed linear span of $S$.
Let $\cC(\cH)$ be the set of all closed linear subspaces in
$\cH$. 
With the set inclusion ordering, 
the set $\cC(\cH)$ is a complete
lattice. 
The operation $M\mapsto M^\perp$ 
is  an orthocomplementation
on the lattice $\cC(\cH)$, with which $\cC(\cH)$ is a
complete orthomodular lattice.
Denote by $\cB(\cH)$ the algebra of bounded linear
operators on $\cH$ and $\cQ(\cH)$ the set of projections on $\cH$.
For any $M\in\cC(\cH)$,
denote by $\cP (M)\in\cQ(\cH)$ the projection operator 
of $\cH$ onto $M$.
Then, 
$M\le N$ if and only if $\cP (M)\subseteq\cP (N)$
for any $M,N\in\cC(\cH)$,
and $\cQ(\cH)$ with the operator ordering is a complete 
orhtomodular lattice isomorphic to $\cC(\cH)$.

Let $\cA\subseteq\cB(\cH)$.
We denote by $\cA'$ the {\em commutant of 
$\cA$ in $\cB(\cH)$}.
A self-adjoint subalgebra $\cM$ of $\cB(\cH)$ is called a
{\em von Neumann algebra} on $\cH$ iff 
$\cM''=\cM$.
We denote by $\cP(\cM)$ the set of projections in
a von Neumann algebra $\cM$.
For any $P,Q\in\cQ(\cH)$, we have 
$P\commutes Q$ iff $[P,Q]=0$, where $[P,Q]=PQ-QP$.
For any subset $\cA\subseteq\cQ(\cH)$,
we denote by $\cA^{!}$ the {\em commutant} 
of $\cA$ in $\cQ(\cH)$.
A {\em logic} on $\cH$ is a sublogic of $\cQ(\cH)$. 
For any subset $\cA\subseteq\cQ(\cH)$, the smallest 
logic including $\cA$ is the logic
$\cA^{!!}$ called the  {\em logic generated by
$\cA$}.  
Then, 
a subset $\cQ \subseteq\cQ(\cH)$ is a logic on $\cH$ if
and only if $\cQ=\cP(\cM)$ for some von Neumann algebra
$\cM$ on $\cH$ \cite{07TPQ}.

We define the {\em implication} and the 
{\em logical equivalence} on $\cQ$ by  
$P\Then Q=P^{\perp}\Or(P\And Q)$
 and
$P\Iff Q=(P\Then Q)\And(Q\Then P)$.
We have the following characterization of commutators in 
logics on Hilbert spaces \cite{07TPQ}.

\begin{Theorem}\label{th:com}
Let $\cQ$ be a logic on $\cH$.
For any subset $\cA\subseteq\cQ$, we have
\[
\com(\cA)=\cP\{\ps\in\cH\mid
[A,B]\ps=0
\mb{ \rm  for all }A,B\in\cA''\}.
\]
\end{Theorem}

\section{Quantum set theory}
\label{se:UQ}

We denote by $\V$ the universe 
of the Zermelo-Fraenkel set theory
with the axiom of choice (ZFC).
Let $\cL(\in)$ be the language 
for first-order theory with equality 
having a binary relation symbol
$\in$, bounded quantifier symbols $\forall x\in y$,
$\exists x \in y$, and no constant symbols.
For any class $U$, 
the language $\cL(\in,U)$ is the one
obtained by adding a name for each element of $U$.

Let $\cQ$ be a logic on $\cH$.
For each ordinal $ {\al}$, let
\[
\V_{\al}^{(\cQ)} = \{u|\ u:\dom(u)\to \cQ \mbox{ and }
(\exists \be<\al)
\dom(u) \subseteq V_{\be}^{(\cQ)}\}.
\]
The {\em $\cQ$-valued universe} $\VL$ is defined
by 
\[
  \VL= \bigcup _{{\al}{\in}\mbox{On}} V_{{\al}}^{(\cQ)},
\]
where $\mbox{On}$ is the class of all ordinals. 
For every $u\in\VQ$, the rank of $u$, denoted by
$\rank(u)$,  is defined as the least $\al$ such that
$u\in \VQ_{\al+1}$.
It is easy to see that if $u\in\dom(v)$ then 
$\rank(u)<\rank(v)$.

For any $u,v\in\VL$, the $\cQ$-valued truth values of
atomic formulas $u=v$ and $u\in v$ are assigned by
by the following rules recursive in rank.
\begin{enumerate}[(i)]\itemsep=0in
\item $\vval{u = v}
= \inf_{u' \in  \cD(u)}(u(u') \Then
\vval{u'  \in v})
\And \inf_{v' \in   \cD(v)}(v(v') 
\Then \vval{v'  \in u})$.
\item $ \vval{u \in v} 
= \sup_{v' \in \cD(v)} (v(v')\And \vval{u =v'})$.
\end{enumerate}

To each statement $\ph$ of $\cL(\in,\VL)$ 
we assign the
$\cQ$-valued truth value $ \val{\ph}_{\cQ}$ by the following
rules.
\begin{enumerate}[(i)]\itemsep=0in
\setcounter{enumi}{2}
\item $ \vval{\Not\ph} = \vval{\ph}^{\perp}$.
\item $ \vval{\ph_1\And\ph_2} 
= \vval{\ph_{1}} \And \vval{\ph_{2}}$.
\item $ \vval{\ph_1\Or\ph_2} 
= \vval{\ph_{1}} \Or \vval{\ph_{2}}$.
\item $ \vval{\ph_1\Then\ph_2} 
= \vval{\ph_{1}} \Then \vval{\ph_{2}}$.
\item $ \vval{\ph_1\Iff\ph_2} 
= \vval{\ph_{1}} \Iff \vval{\ph_{2}}$.
\item $ \vval{(\forall x\in u)\, {\ph}(x)} 
= \Inf_{u'\in \dom(u)}
(u(u') \Then \vval{\ph(u')})$.
\item $ \vval{(\exists x\in u)\, {\ph}(x)} 
= \Sup_{u'\in \dom(u)}
(u(u') \And \vval{\ph(u')})$.
\item $ \vval{(\forall x)\, {\ph}(x)} 
= \Inf_{u\in \VL}\vval{\ph(u)}$.
\item $ \vval{(\exists x)\, {\ph}(x)} 
= \Sup_{u\in \VL}\vval{\ph(u)}$.
\end{enumerate}

We say that a statement ${\ph}$ of $ \cL(\in,\VL) $
{\em holds} in $\VL$ if $ \val{{\ph}}_{\cQ} = 1$.
A formula in $\cL(\in)$ is called a {\em
$\De_{0}$-formula}  if it has no unbounded quantifiers
$\forall x$ or $\exists x$.
The following theorem holds \cite{07TPQ}.

\sloppy
\begin{Theorem}[$\De_{0}$-Absoluteness Principle]
\label{th:Absoluteness}
\sloppy  
For any $\De_{0}$-formula 
${\ph} (x_{1},{\ldots}, x_{n}) $ 
of $\cL(\in)$ and $u_{1},{\ldots}, u_{n}\in \VQ$, 
we have
\[
\val{\ph(u_{1},\ldots,u_{n})}_{\cQ}=
\val{\ph(u_{1},\ldots,u_{n})}_{\cQ(\cH)}.
\]
\end{Theorem}
Henceforth, 
for any $\De_{0}$-formula 
${\ph} (x_{1},{\ldots}, x_{n}) $
and $u_1,\ldots,u_n\in\VQ$,
we abbreviate $\val{\ph(u_{1},\ldots,u_{n})}=
\val{\ph(u_{1},\ldots,u_{n})}_{\cQ}$,
which is the common $\cQ$-valued truth value 
in all $\VL$ such  that $u_{1},\ldots,u_{n}\in\VL$.

The universe $\V$  can be embedded in
$\VQ$ by the following operation 
$\vee:v\mapsto\check{v}$ 
defined by the $\in$-recursion: 
for each $v\in\V$, $\check{v} = \{\check{u}|\ u\in v\} 
\times \{1\}$. 
Then we have the following \cite{07TPQ}.
\begin{Theorem}[$\De_0$-Elementary Equivalence Principle]
\label{th:2.3.2}
\sloppy  
For any $\De_{0}$-formula 
${\ph} (x_{1},{\ldots}, x_{n}) $ 
of $\cL(\in)$ and $u_{1},{\ldots}, u_{n}\in V$,
we have
$
\bracket{\V,\in}\models  {\ph}(u_{1},{\ldots},u_{n})
\mbox{ if and only if }
\val{\ph(\check{u}_{1},\ldots,\check{u}_{n})}=1.
$
\end{Theorem}

For $u\in\VQ$, we define the {\em support} 
of $u$, denoted by $L(u)$, by transfinite recursion on the 
rank of $u$ by the relation
\[
L(u)=\bigcup_{x\in\dom(u)}L(x)\cup\{u(x)\mid x\in\dom(u)\}.
\]
For $\cA\subseteq\VQ$ we write 
$L(\cA)=\bigcup_{u\in\cA}L(u)$ and
for $u_1,\ldots,u_n\in\VQ$ we write 
$L(u_1,\ldots,u_n)=L(\{u_1,\ldots,u_n\})$.
Let $\cA\subseteq\VQ$.  The {\em commutator
of $\cA$}, denoted by $\com(\cA)$, is defined by 
\[
\cuniv(\cA)=\com (L(\cA)).
\]
For any $u_1,\ldots,u_n\in\VQ$, we write
$\cuniv(u_1,\ldots,u_n)=\cuniv(\{u_1,\ldots,u_n\})$.
For bounded theorems,  the following transfer 
principle holds \cite{07TPQ}.

\begin{Theorem}[$\De_{0}$-ZFC  Transfer Principle]
For any $\De_{0}$-formula ${\ph} (x_{1},{\ldots}, x_{n})$ 
of $\cL(\in)$ and $u_{1},{\ldots}, u_{n}\in\VQ$, if 
${\ph} (x_{1},{\ldots}, x_{n})$ is provable in ZFC, then
we have
\[
\cuniv(u_{1},\ldots,u_{n})\le
\val{\ph({u}_{1},\ldots,{u}_{n})}.
\]
\end{Theorem}

\section{Real numbers in quantum set theory}
\label{se:RN}

Let $\Q$ be the set of rational numbers in $V$.
We define the set of rational numbers in the model $\VQ$
to be $\check{\Q}$.
We define a real number in the model by a Dedekind cut
of the rational numbers. More precisely, we identify
a real number with the upper segment of a Dedekind cut
assuming that the lower segment has no end point.
Therefore, the formal definition of  the predicate $\R(x)$, 
``$x$ is a real number,'' is expressed by
\[
\R(x):=
\forall y\in x(y\in\check{\Q})
\And \exists y\in\check{\Q}(y\in
x)
\And \exists y\in\check{\Q}(y\not\in x)\And
\forall y\in\check{\Q}(y\in x\Iff\forall z\in\check{\Q}
(y<z \Then z\in x)).
\]
The symbol ``:='' is used to define a new formula, here and hereafter.
We define $\R^{(\cQ)}$ to be the interpretation of 
the set $\R$ of real
numbers in $\VQ$ as follows.
\[
\R^{(\cQ)} = \{u\in\VQ|\ \cD(u)=\cD(\check{\Q})
\ \mb{and }\val{\R(u)}=1\}.
\]
The set $\R_\cQ$ of real numbers in $\VQ$ is defined by
\[
\R_\cQ=\R^{(\cQ)}\times\{1\}.
\]
For any $u,v\in\R^{(\cQ)}$, 
Then, the following relations hold in $\VQ$ \cite{07TPQ}.
\begin{enumerate}[(i)]\itemsep=0in
\item $\val{(\forall u\in\R_\cQ) u=u}=1$.
\item $\val{(\forall u,v\in\R_\cQ) u=v\Then v=u}=1$.
\item $\val{(\forall u,v,w\in\R_\cQ) u=v\And v=w\Then u=w}=1$.
\item $\val{(\forall v\in\R_\cQ)(\forall x,y\in v)x=y\And x\in v\Then
y\in v}$.
\item $\val{(\forall u,v\in\R_\cQ)(\forall x\in u)x\in u\And u=v\Then x\in
v}.$
\end{enumerate}
From the above, 
the equality is an equivalence relation between real numbers in $\VQ$.
%
%
%
For any $u_1,\ldots,u_n\in\R^{(\cQ)}$, we have
\[
\val{u_1=u_2\And\cdots\And u_{n-1}=u_n}\le 
\cuniv(u_1,\ldots,u_n),
\]
and hence commutativity follows from equality in $\R^{(\cQ)}$ \cite{07TPQ}.

Let $\cM$ be a von Neumann algebra on a Hilbert
space $\cH$ and let $\cQ=\cP(\cM)$.
A closed operator $A$ (densely defined) on $\cH$ is
said to be {\em affiliated} with $\cM$, in symbols $A\,\et\,\cM$, 
iff $U^{*}AU=A$ for any unitary operator $U\in\cM'$.
Let $A$ be a self-adjoint operator (densely defined) on $\cH$
and let $A=\int_{\R} \la\, dE^A(\la)$
be its spectral decomposition, where $\{E^{A}(\la)\}_{\la\in\R}$ 
is the resolution of  identity belonging to $A$ \cite[p.\ 119]{vN55}.
It is well-known that $A\af\cM$ if and only if $E^A({\la})\in\cQ$ 
for every $\la\in\R$.
Denote by $\overline{\cM}_{SA}$ the set of self-adjoint operators 
affiliated with $\cM$.
Two self-adjoint operators $A$ and $B$ are said to {\em commute},
in symbols $A\commutes B$,
iff $E^A(\la)\commutes E^B(\la')$ for every pair 
$\la,\la'$ of reals.

Let $\cB$ be a Boolean logic on $\cH$. 
For any $u\in\R^{(\cB)}$ and $\la\in\R$, we define $E^{u}(\la)$ by 
\[
E^{u}(\la)=\Inf_{\la<r\in\Q}u(\check{r}).
\]
Then, it can be shown that
$\{E^u(\la)\}_{\la\in\R}$ is a resolution of
identity in $\cB$ and hence by the spectral theorem there
is a self-ajoint operator $\hat{u}\af\cB''$ uniquely
satisfying $\hat{u}=\int_{\R}\la\, dE^u(\la)$.  On the other
hand, let $A\af\cB''$ be a self-ajoint operator. We define $\tilde{A}\in\VB$ by
\[
\dom(\tilde{A})=\dom(\check{\Q})\mbox{ and }
\tilde{A}(\check{r})=E^{A}(r)\mbox{ for all }r\in\Q.
\]
Then, it is easy to see that $\tilde{A}\in\RB$ and we have
$(\hat{u})\tilde{}=u$ for all $u\in\RB$ and $(\tilde{A})\hat{}=A$
for all $A\in\overline{(\cB'')}_{SA}$.
Therefore, the correspondence
between $\RB$ and $\overline{(\cB'')}_{SA}$ is a one-to-one correspondence.
We call the above correspondence the {\em Takeuti correspondence}.
Now, we have the following \cite{07TPQ}.

\begin{Theorem}
Let $\cQ$ be a logic on $\cH$.  The relations 
\begin{enumerate}[(i)]\itemsep=0in
\item ${\displaystyle E^{A}(\la)=\Inf_{\la<r\in\Q}u(\check{r})}$ for all $\la\in\Q$,
\item $u(\check{r})=E^{A}(r)$ for all $r\in\Q$,
\end{enumerate}
for all $u=\tA\in\RQ$ and $A=\hu\in \overline{(\cQ'')}_{SA}$ 
sets up a one-to-one correspondence between $\RQ$ and $ \overline{(\cQ'')}_{SA}$.
\end{Theorem}

\section{Standard probabilistic interpretation} 
\label{se:2}

Let $\cH$ be a Hilbert space describing a quantum system $\bS$.
For the system $\bS$, the {\em observables} are defined as self-adjoint operators on $\cH$, 
the {\em states} are defined as density operators, and a {\em vector state} $\psi$ is identified
with the state $\ketbra{\psi}$.
We denote by $\cO(\cH)$ the set of observables,
by $\cS(\cH)$ the space of density operators, and 
by $\cB(\cH)$ the space of bounded operators on $\cH$.
Observables $X_1,\ldots,X_n\in\cO(\cH)$ are said to be {\em mutually commuting} if
$X_j\commutes X_k$ for all $j,k=1,\ldots,n$.
If $X_1,\ldots,X_n\in\cO(\cH)$ are bounded, this condition is equivalent to 
$[X_j,X_k]=0$ for all $j,k=1,\ldots,n$.
The standard probabilistic interpretation of quantum theory defines the 
{\em joint probability distribution function} $F^{X_1,\ldots,X_n}_{\rh}(x_1,\ldots,x_n)$
for mutually commuting observables $X_1,\ldots,X_n\in\cO(\cH)$ in 
$\rh\in\cS(\cH)$ by the {\em Born statistical formula}:
\[
F^{X_1,\ldots,X_n}_{\rh}(x_1,\ldots,x_n)=\Tr[E^{X_1}(x_1)\cdots E^{X_n}(x_n)\rh].
\]

To clarify the logical structure presupposed in the standard probabilistic interpretation,
we define {\em observational propositions} for $\bS$ 
by the following rules.
\begin{itemize}\itemsep=0in
\item[(R1)] For any $X\in\cO(\cH)$ and $x\in \R$, the expression
$X\le x$ is an observational proposition.
\item[(R2)] If $\ph_1$ and $\ph_2$ are observational propositions,
$\Not \ph_1$ and $\ph_1\And \ph_2$ are also observational propositions.
\end{itemize}
Thus, every observational proposition is built up from ``atomic'' 
observational propositions $X\le x$ by means of the connectives 
$\Not$ and $\And$.
We introduce the connective $\Or$ by definition.
\bitem
\item[(D1)] $\ph_1\Or\ph_2:= \Not(\Not \ph_1\And\Not\ph_2)$.
\eitem

For each observational proposition $\ph$, we assign its 
projection-valued truth value $\valo{\ph}\in\cQ(\cH)$ 
by the following rules \cite{BvN36}.
\bitem
\item[(T1)] $\valo{X\le x}=E^{X}(x).$
\item[(T2)] $\valo{\Not \ph}=\valo{\ph}^{\perp}.$
\item[(T3)] $\valo{\ph_1\And \ph_2}=\valo{\ph_1}\And\valo{\ph_2}.$
\eitem
From (D1),  (T2) and (T3), we have
\bitem
\item[(D2)] $\valo{\ph_1\Or \ph_2}=\valo{\ph_1}\Or\valo{\ph_2}.$
\eitem

We define the {\em probability} $\Pr\{\ph\|\rh\}$ 
of an observational proposition $\ph$ in a state $\rh$ by
\bitem
\item[(P1)] $\Pr\{\ph\|\rh\}=\Tr[\valo{\ph}\rh]$.
\eitem
We say that {\em an observational proposition $\ph$ holds in a state $\rh$} if
$\Pr\{\ph\|\rh\}=1$.

The standard interpretation of quantum theory restricts observational propositions
to be standard defined as follows.
\bitem 
\item[(W1)] An observational proposition including atomic formulas
$X_1\le x_1, \ldots,X_n\le x_n$ is called {\em standard} if $X_1,\ldots,X_n$
are mutually commuting.   
\eitem 

All the standard observational propositions including only given mutually commuting observables 
$X_1, \ldots,X_n$ comprise a complete Boolean algebra under the logical order $\le$ defined by
$\ph\le \ph'$ iff $\valo{\ph}\le \valo{\ph'}$ and obey inference rules in classical logic.
Suppose that $X_1,\ldots,X_n\in\cO(\cH)$ are mutually commuting.
Let $x_1,\ldots,x_n\in\R$.
Then, 
$X_1\le x_1\And\cdots\And X_n\le x_n$ is a standard observational proposition.
We have
\[
\valo{ X_1\le x_1\And\cdots\And X_n\le x_n}
=E^{X_1}(x_1)\And\cdots\And E^{X_n}(x_n)
=E^{X_1}(x_1)\cdots E^{X_n}(x_n).
\]
Hence, we reproduce the Born statistical formula as 
\[
\Pr\{X_1\le x_1\And\cdots\And X_n\le x_n\|\rh\}
=\Tr[E^{X_1}(x_1)\cdots E^{X_n}(x_n)\rh].
\]
From the above, our definition of the truth vales of observational propositions 
are consistent with the standard probabilistic interpretation of quantum theory. 

In order to make the counter part of $r\in\R$ in $\RQ$,  
for any $r\in \R$, we define $\tilde{r}\in\RQ$ by
\beqas
\cD(\tilde{r})=\cD(\check{\Q})\quad\mbox{and}\quad
\tilde{r}(\check{t})=\val{\check{r}\le \check{t}}
\eeqas
for all $t\in\Q$.
Then, $\tilde{r}\in\RQ$ corresponds to $\tilde{(r1)}$, where $1$ is the identity operator, 
under the Takeuti correspondence.

For every observational proposition $\ph$ the corresponding statement 
$\tilde{\ph}$ in $\cL(\in,\RQ)$ is given by the following rules for any $X\in\cO(\cH)$,
$x\in\R$, and observational propositions $\ph,\ph_1,\ph_2$.
\bitem
\item[(Q1)] $\displaystyle\widetilde{X\le x}:=\tilde{X}\le\tilde{x}$.
\item[(Q2)] $\widetilde{\Not \ph}:=\Not\tilde{\ph}.$
\item[(Q3)] $\widetilde{\ph_1\And \ph_2}:={\tilde\ph_1}\And\tilde{\ph_2}.$
\eitem
Then, it is easy to see that the relation
$$
\val{\tilde{\ph}}=\valo{\ph}
$$
holds for any observational proposition $\ph$.  
Thus, all the observational propositions are embedded in statements in $\cL(\in,\RQ)$
with the same projection-valued truth value assignments.

Let $E^{X}(\la)$ be the resolution of identity belonging 
to $X\in\cO(\cH)$.
Let $a<b\in\R$.  For the interval $I=(a,b]$, we define 
$
E^{X}(I)=E^{X}(b)-E^{X}(a),
$
and we define the corresponding interval 
$\tilde{I}$ of real numbers in $\VQ$ by
$
\cD(\tilde{I})=\RQ$
and
$\tilde{I}(u)=\val{ u\le \tilde{a}}^{\perp}\And\val{u\le\tilde{b}}$
for all $u\in\RQ$.   Then, we have
$
\val{\tilde{X}\in\tilde{I}}=E^{X}(I). 
$
The observational proposition $X\in I$,
which will be also denoted by $a<X\le b$, is defined as
\[
X\in I:=\Not(a\le X)\And(X\le b).
\]
Then, we have $\valo{X\in I}=\val{\tilde{X}\in\tilde{I}}$.
For mutually commuting observables $X_1,\ldots,X_n\in\cO(\cH)$ and
intervals $I_{1}=(a_1,b_1],\ldots,I_{n}=(a_n.b_n]$ we have
\[
\Pr\{X_1\in I_1\And\cdots\And X_n\in I_n\|\rh\}
=\Tr[E^{X_1}(I_1)\cdots E^{X_n}(I_n)\rh].
\]

\section{Joint determinateness}
\label{se:3}

Let $\cO_{\om}(\cH)$ be the set of observables on $\cH$ with finite spectra.
An observable $X\in\cO(\cH)$ is said to be {\em finite} if $X\in \cO_{\om}(\cH)$,
and {\em infinite} otherwise.
Let $X\in\cO_{\om}(\cH)$.  Let 
$
\de(X)=\min_{x,y\in\Sp(X), x\not= y}\{|x-y|/2, 1\}.
$
For any $x\in\R$,  we define the formula $X=x$ by
\[
X=x:=x-\de(X)<X\le x+\de(X).
\] 
Then, it is easy to see that we have  
\[
\valo{X=x}=\cP\{\psi\in\cH\mid X\psi=x\psi\}
\]
for all $x\in\R$.

\sloppy
For observational propositions $\ph_1,\ldots,\ph_n$, we define the
observational proposition $\Sup_{j}\ph_j$ by
$\Sup_{j}\ph_j=\ph_1\Or\cdots\Or\ph_n$.
We denote by $\Sp(X)$ the spectrum of an observable $X\in\cO(\cH)$.
For any finite observables
$X_1,\ldots,X_n\in\cO_{\om}(\cH)$ we define the observational proposition
$\cm(X_1,\ldots,X_n)$ by
\[
\cm(X_1,\ldots,X_n):=
\Sup_{x_1\in\Sp(X_1),\ldots,x_n\in\Sp(X_n)} 
X_1=x_1\And\cdots\And X_n=x_n.
\]
We say that observables $X_1,\ldots,X_n$ are {\em jointly determinate} 
in a state $\rh$ if the observational proposition
$\cm(X_1,\ldots,X_n)$  holds in $\rh$.
In general, we say that observables $X_1,\ldots,X_n$ are {\em jointly determinate} 
in a state $\rh$ with probability $\Pr\{\cm(X_1,\ldots,X_n)\|\rh\}$.
Then, we have the following \cite{11QRM}.
\begin{Theorem}\label{th:SD}
Finite observables  $X_1,\ldots,X_n\in\cO_{\om}(\cH)$ are jointly determinate in
a vector state $\ps$  if and only if the state $\ps$ is a superposition
of common eigenvectors of $X_1,\ldots,X_n$.
\end{Theorem}

The joint determinateness is characterized by the commutator in quantum set
theory as follows.
\begin{Theorem}For any finite observables  $X_1,\ldots,X_n\in\cO_{\om}(\cH)$,
we have
\beql{JD}
\valo{\cm(X_1,\ldots,X_n)}=\com(\tX_1,\ldots,\tX_n).
\eeq
\end{Theorem}

For self-adjoint operators $A_1,\ldots,A_n$ on $\cH$, 
the {\em von Neumann algebra generated by $A_1,\ldots,A_n$}, denoted by
$\{A_1,\ldots,A_n\}''$, is  the von Neumann algebra generated by projections 
$E^{A_{j}}(x)$ for all $j=1,\ldots,n$ and $x\in\R$. 
Under the Takeuti correspondence, the commutator of quantum reals are characterized 
as follows.
\begin{Theorem}\label{th:com_real}
Let $\cQ$ be a logic on $\cH$ and
let $u_1,\ldots,u_n\in\R^{(\cQ)}$. 
Then we have
\[
\com(u_1,\ldots,u_n)=\cP\{\ps\in\cH\mid
[A,B]\ps=0
\mb{ \rm  for all }A,B\in\{\hu_1,\ldots,\hu_n\}''\}.
\]
\end{Theorem}

Although we cannot find an observational proposition $\cm(X_1,\ldots,X_n)$ satisfying
\Eq{JD} for infinite observables $X_1,\ldots,X_n\in\cO(\cH)$, we can introduce a new
atomic observational propositions $ \cm(X_1,\ldots,X_n)$  with \Eq{JD} for all
 $X_1,\ldots,X_n\in\cO(\cH)$.  
We introduce the following additional rule for formation of observational
 propositions:
 \bitem
 \item[(R3)] For any $X_1,\ldots,X_n\in\cO(\cH)$ and $x_1,\ldots,x_n\in \R$, the expression
$\cm(X_1,\ldots,X_n)$ is an observational proposition.
 \eitem
 Moreover, we introduce the following additional rule for  projection-valued truth values:
 \bitem 
\item[(T4)]
$ \valo{\cm(X_1,\ldots,X_n)}=\cP\{\ps\in\cH\mid
[A,B]\ps=0
\mb{ \rm  for all }A,B\in\{\hu_1,\ldots,\hu_n\}''\}.$
 \eitem
From Theorem \ref{th:com_real}, \Eq{JD} holds for any 
 $X_1,\ldots,X_n\in\cO(\cH)$ under (T4).
Thus, 
we naturally extend the notion of joint determinateness to arbitrary observables.
We say that observables $X_1,\ldots,X_n\in\cO(\cH)$ are {\em jointly determinate} in
a state $\rh$ if $\Pr\{\cm(X_1,\ldots,X_n)\|\rh\}=1$, or equivalently if
$\Tr[\com(X_1,\ldots,X_n)\rh]=1$.  It is easy to see that this condition is equivalent 
to that $[A,B]\rh=0$ for all $A,B\in\{\tX_1,\ldots,\tX_n\}''$.

\sloppy
A probability distribution function $F(x_1,\ldots,x_n)$ on $\R^{n}$,
is called a {\em joint probability distribution function} of 
$X_1,\ldots,X_n\in\cO(\cH)$ in $\rh\in\cS(\cH)$ if
\[
F(x_1,\ldots,x_n)
=\Pr\{X_1\le x_1\And\cdots\And X_n\le x_n\|\rh\}.
\]
A joint probability distribution $F$ of $X_1,\ldots,X_n$ in $\rh$ is unique, if any, 
and denoted by $F^{X_1,\cdots X_n}_{\rh}(x_1,\ldots,x_n)$.

Since the joint determinateness is considered to be the state-dependent 
notion of commutativity, it is expected that the joint determinateness
is equivalent to the state-dependent existence of the joint probability distribution function,
as shown below.
 
\sloppy
\begin{Theorem}\label{th:JPD}
Observables $X_1,\ldots,X_n\in\cO(\cH)$ are
jointly determinate in a state $\rh$
if and only if
there exists a joint probability distribution function 
$F^{X_1,\cdots X_n}_{\rh}(x_1,\ldots,x_n)$
of $X_1,\ldots,X_n$ in
$\rh$. 
In this case, for any polynomial $p(f_1(X_1),\ldots,f_n(X_n))$ of observables
$f_1(X_1),\ldots,f_n(X_n)$, where $f_1,\ldots,f_n$ are bounded Borel functions, 
we have
\[
\Tr[p(f_1(X_1),\ldots,f_n(X_n))\rh]
=
\idotsint_{\R^{n}}p(f_1(x_1),\ldots,f_n(x_n))\,
F^{X_1,\cdots X_n}_{\rh}(dx_1,\ldots,dx_n).
\]
\end{Theorem}

\section{Quantum equality}
\label{se:4}

For any finite observables $X,Y$, we define 
the observational proposition $X=Y$ by 
\[
X=Y:=
\Sup_{x\in \Sp(X)}X=x\And Y=x.
\]
We say that observables $X$ and $Y$ {\em are equal in a state $\rh$}
if $X=Y$ holds in $\rh$.
In this case, we shall write $X=_\rh Y$.
In general, we say that 
observables $X$ and $Y$ {\em are equal in a state $\rh$
with probability $\Pr\{X=Y\|\rh\}$}.
Then, we have the following \cite{11QRM}.
\begin{Theorem}\label{th:EQ}
Finite observables  $X$ and $Y$ are equal in 
a vector state $\ps$  if and only if the state $\ps$ is a superposition
of common eigenvectors of $X$ and $Y$ with common eigenvalues.
\end{Theorem}

The state-dependent equality is characterized by the equality in quantum set
theory as follows.
\begin{Theorem}For any finite observables  $X,Y\in\cO_{\om}(\cH)$,
we have
\beql{QE}
\valo{X=Y}=\val{\tX=\tY}.
\eeq
\end{Theorem}

Under the Takeuti correspondence, the truth values of equality between 
reals are characterized as follows.
\begin{Theorem}\label{th:eq_real}
Let $\cQ$ be a logic on $\cH$ and
let $u,v\in\R^{(\cQ)}$. 
Then we have
\[
\val{u=v}=\cP\{\psi\in\cH\mid E^{\hat{u}}(\la)\psi= E^{\hat{v}}(\la)\psi
\mb{ \rm  for all }\la\in\R\}.
\]
\end{Theorem}

We cannot find an observational proposition $X=Y$ satisfying
\Eq{QE} for infinite observables $X,Y\in\cO(\cH)$.
We introduce a new atomic observational propositions $X=Y$  with \Eq{QE} for all
 $X,Y\in\cO(\cH)$ by the following additional rules for formation of observational
 propositions and for projection-valued truth values:
 \bitem
 \item[(R4)] For any $X,Y\in\cO(\cH)$ and $x,y\in \R$, the expression
$X=Y$ is an observational proposition.
\item[(T5)]
$ \valo{X=Y}=
\cP\{\psi\in\cH\mid E^{X}(\la)\psi= E^{Y}(\la)\psi
\mb{ \rm  for all }\la\in\R\}.
$
 \eitem
Note that from Theorem \ref{th:eq_real}, \Eq{QE} holds for any 
 $X,Y\in\cO(\cH)$.
We say that observables $X$ and $Y$ are {\em equal} in
a state $\rh$ if $\Pr\{X=Y\|\rh\}=1$, or equivalently if
$\Tr[\valo{X=Y}\rh]=1$.  It is easy to see that this condition is equivalent 
to that $E^{X}(\la)\rho= E^{Y}(\la)\rho$ for all $\la\in\R$.
Thus, we naturally extend the state-dependent notion of equality to arbitrary observables.

\begin{Theorem}\label{th:ID}
For any observables $X,Y\in\cO_{\om}(\cH)$ and $\rh\in\cS(\cH)$, we have 
$X=_\rh Y$ if and only if there exists a joint probability distribution function 
$F^{X,Y}_{\rho}(x,y)$ of $X,Y$ in $\rho$ and it satisfies
\[
\iint_{\De}\,F^{X,Y}_{\rho}(x,y)=1,
\]
where $\De$ is the diagonal set  $\De=\{(x,y)\in\R^{2}\mid x=y\}$. 
\end{Theorem}

Let $\ph(X_1,\ldots,X_n)$ be an observational proposition 
that includes symbols for observables only from the list $X_1,\ldots,X_n$.
Then, $\ph(X_1,\ldots,X_n)$ is said to be {\em contextually well-formed} in a state $\rh$
if $X_1,\ldots,X_n$ are jointly determinate in $\rh$.
The following theorem is an easy consequence from the transfer principle in quantum 
set theory \cite{07TPQ}, and shows that for well-formed observational propositions 
$\ph(X_1,\ldots,X_n)$ for a fixed family $X_1,\ldots,X_n$ of observables,
the projection-valued truth value assignments satisfy Boolean inference rules and
the probability assignments satisfy rules for calculus of classical probability. 

\begin{Theorem}
If $\ph(X_1,\ldots,X_n)$ is a tautology in classical logic,
then we have 
\[
\valo{\cm(X_1,\ldots,X_n)}\le\valo{\ph(X_1,\ldots,X_n)}.
\]
Moreover, if $\ph(X_1,\ldots,X_n)$ is contextually well-formed
in a state $\rh$, then $\ph(X_1,\ldots,X_n)$ holds in $\rh$.
\end{Theorem}

\section{Measurements of observables}
\label{se:6}

A {\em measuring process} for $\cH$ is defined to be a quadruple
$(\cK,\si,U,M)$ consisting of a Hilbert space $\cK$, 
a state $\si$ on $\cK$, 
a unitary operator $U$ on $\cH\otimes\cK$,
and an observable $M$ on $\cK$ \cite{84QC}.
A measuring process $\M(\bx)=(\cK,\si,U,M)$ with {\em output variable} $\bx$
describes a
measurement carried out by an interaction, called the {\em measuring interaction}, 
from time 0 to time $\De t$ between the measured system
$\bS$ described by $\cH$ and the {\em probe} system $\bP$ described by $\cK$ 
that is prepared in the state $\si$ at time 0.
The outcome of this measurement is obtained by
measuring the observable $M$, called the {\em meter observable},  
in the probe at time $\De t$.
The unitary operator $U$ describes the time evolution of $\bS+\bP$ from time 0 to
$\De t$.  We shall write $M(0)=1\otimes M$, $M(\De t)=U^{\da}M(0)U$, 
$X(0)=X\otimes 1$,  and $X(\De t)=U^{\da}X(0)U$
for any observable $X\in\cO(\cH)$.
We can use the probabilistic interpretation for the system $\bS+\bP$.
The {\em output distribution}
$\Pr\{\bx\le x\|\rho\}$,
the probability distribution function 
of the output variable $\bx$ of this measurement on input state $\rh\in\cS(\cH)$, 
is naturally defined as
\[
\Pr\{\bx\le x\|\rho\}=\Pr\{M(\De t)\le x\|\rho\otimes\si\}=\Tr[E^{M(\De t)}(x)\rho\otimes\si].
\]
The POVM of  the measuring process $\M(\bx)$ is defined by
\[
\Pi(x)
=\Tr_{\cK}[E^{M(\De t)}(x)(I\otimes\si)].
\]
Then, we have 
\bitem
\item[(P1)] $\lim_{x\to -\infty}\Pi(x)=0$,  $\lim_{x\to +\infty}\Pi(x)=1$, and $\lim_{x_0\le x\to x_0}\Pi(x)=\Pi(x_0)$,
\item[(P2)] $\Pi(x')\le \Pi(x'')$ for $x'\le x''$,
\item[(P3)] $\Pr\{\bx\le x\|\rh\}=\Tr[\Pi(x)\rho]$.
\eitem
Conversely, it has been proved in Ref.~\cite{84QC} that
for every $\{\Pi(x)\}_{x\in\R}$ satisfying (P1), and (P2),
there is a measuring process $(\cK,\si,U,M)$ satisfying (P3).

Let $A\in\cO(\cH)$ and $\rh\in\cS(\cH)$.
A measuring process $\M(\bx)=(\cK,\si,U,M)$ with the POVM $\Pi(x)$
is said to {\em measure $A$} in $\rho$
if $A(0)=_{\rho\otimes\si}M(\De t)$,
and {\em weakly measure} $A$ in $\rho$ if
$
\Tr[\Pi(x)E^{A}(y)\rho]=\Tr[E^{A}(\min\{x,y\})\rho]
$
for any $x,y\in\R$.
A measuring process $\M(\bx)$ is said to 
{\em satisfy the Born statistical formula} (BSF)  for $A$
in $\rho$ if it satisfies
$
\Pr\{\bx\le x\|\rho\}=\Tr[E^{A}(x)\rho]
$
for all $x\in\R$.
The following theorem characterizes measurements 
of an observable in a given state.

\begin{Theorem}
Let $\M(\bx)=(\cK,\si,U,M)$ be a measuring process for $\cH$ with the POVM $\Pi(x)$.
For any observable $A\in\cO(\cH)$ and any state $\rh\in\cS(\cH)$,
the following conditions are all equivalent.
\begin{enumerate}[{\rm (i)}]\itemsep=0in
\item $\M(\bx)$ measures $A$ in $\rho$.
\item $\M(\bx)$ weakly measures $A$ in $\rho$.
\item $\M(\bx)$ satisfies the BSF for $A$ in any vector state
 $\psi\in\cC(A,\rho)$.
\eenum
\end{Theorem}

\begin{Theorem}
Let $\M(\bx)=(\cK,\si,U,M)$ be a measuring process for $\cH$ with the POVM $\Pi(x)$.
Then, $\M(\bx)$ measures $A\in\cO(\cH)$ in any $\rh\in\cS(\cH)$
if and only if\/ $\Pi(x)=E^{A}(x)$ for all $x\in\R$.
\end{Theorem}

\section{Simultaneous measurability}
\label{se:7}

For any measuring process $\M(\bx)=(\cK,\si,U,M)$ and a real-valued Borel function $f$, 
the measuring process $\M(f(\bx))$ with output variable 
$f(\bx)$ is defined by $\M(f(\bx))=(\cK,\si,U,f(M))$.  Observables $A,B$ are 
said to be {\em simultaneously measurable} in a state $\rh\in\cS(\cH)$
by $\M(\bx)$ if there are Borel functions $f,g$ such that
$\M(f(\bx))$ and $\M(g(\bx))$ measure $A$ and $B$ in $\rh$,
respectively.
Observables $A,B$ are said to be {\em simultaneously measurable}
 in $\rh$ if there is a measuring process $\M(\bx)$ such that 
$A$ and $B$ are simultaneously measurable in $\rh$ by $\M(\bx)$.
 
The cyclic subspace $\cC(A,B,\rho)$ of $\cH$ generated by $A,B$ and $\rho$ is
defined by 
\[
\cC(A,B,\rho)=(\{A,B\}''\rh \cH)^{\perp\perp}.
\]
We define $\cC(A,\rho)=\cC(A,1,\rho)$ and $\cC(B,\rho)=\cC(1,B,\rho)$.

The simultaneous measurability and the commutativity are not
equivalent notion under the state-dependent formulation, as
the following theorem clarifies.

\begin{Theorem}\label{th:main}
{\rm (i)} Two observables $A,B\in\cO(\cH)$ are 
jointly determinate in a state $\rh\in\cS(\cH)$ 
if and only if there exists a POVM $\Pi(x,y)$ on $\R^{2}$ satisfying
\beqas
\lim_{y\to +\infty}\Pi(x,y)&=&E^{A}(x)\quad\mbox{on $\cC(A,B,\rho)$ for all $x\in\R$},\\
\lim_{x\to +\infty}\Pi(x,y)&=&E^{B}(y)\quad\mbox{on $\cC(A,B,\rho)$ for all $y\in\R$}.
\eeqas

{\rm (ii)} Two observables $A,B\in\cO(\cH)$ are
simultaneously measurable in a state $\rh\in\cS(\cH)$
if and only if there exists a POVM $\Pi(x,y)$ on $\R^{2}$ satisfying
\beqas
\lim_{y\to +\infty}\Pi(x,y)=E^{A}(x)\quad\mbox{on $\cC(A,\rho)$ for all $x\in\R$}, \\
\lim_{x\to +\infty}\Pi(x,y)=E^{B}(y)\quad\mbox{on $\cC(B,\rho)$ for all $y\in\R$}.
\eeqas
\end{Theorem}

\section{Conclusion}
To formulate the standard probabilistic interpretation of quantum theory,
we have introduced the language of observational propositions with rules (R1) and (R2) 
for well-formed formulas constructed from atomic formulas of the form $X\le x$, 
rules  (T1), (T2), and (T3) for projection-valued truth 
value assignment, and rule (P1) for probability assignment.
Then, the standard probabilistic interpretation gives the statistical predictions
for standard observational propositions formulated by (W1), 
which concern only commuting family of observables.
The Born statistical formula is naturally derived in this way.
We have extended the standard interpretation by introducing two types of atomic
formulas $\com(X_1,\ldots,X_n)$ for joint determinateness and 
$X=Y$ for equality.  To extended observational propositions 
formed through rules (R1), $\ldots$, (R4),  the projection-valued 
truth values are assigned by rule (T1), $\ldots$, (T5), and the probability
assignments are given by rule (P1).  
Then, we can naturally extend the standard interpretation to a general
and state-dependent interpretation for observational propositions including
the relations of joint determinateness and equality.
Quantum set theory ensures that any contextually well-formed formula
provable in ZFC has probability assignment to be 1.
This extends the classical inference for quantum theoretical predictions 
from commuting observables to jointly determinate observables.
We apply this new interpretation to construct a theory of measurement of 
observables and a theory of simultaneous measurement in the state-dependent
approach, to which the standard interpretation cannot apply.
We have reported only basic formulations here, but further development 
in this approach will be reported elsewhere.


\begin{thebibliography}{10}
\providecommand{\bibitemdeclare}[2]{}
\providecommand{\surnamestart}{}
\providecommand{\surnameend}{}
\providecommand{\urlprefix}{Available at }
\providecommand{\url}[1]{\texttt{#1}}
\providecommand{\href}[2]{\texttt{#2}}
\providecommand{\urlalt}[2]{\href{#1}{#2}}
\providecommand{\doi}[1]{doi:\urlalt{http://dx.doi.org/#1}{#1}}
\providecommand{\bibinfo}[2]{#2}

\bibitemdeclare{article}{BvN36}
\bibitem{BvN36}
\bibinfo{author}{G.~\surnamestart Birkhoff\surnameend} \&
  \bibinfo{author}{J.~\surnamestart von Neumann\surnameend}
  (\bibinfo{year}{1936}): \emph{\bibinfo{title}{The Logic of Quantum
  Mechanics}}.
\newblock {\sl \bibinfo{journal}{Ann.\ Math.}} \bibinfo{volume}{{37}}, pp.
  \bibinfo{pages}{823--845}, \doi{10.2307/1968621}.

\bibitemdeclare{article}{Che89}
\bibitem{Che89}
\bibinfo{author}{G.~\surnamestart Chevalier\surnameend} (\bibinfo{year}{1989}):
  \emph{\bibinfo{title}{Commutators and Decompositions of Orthomodular
  Lattices}}.
\newblock {\sl \bibinfo{journal}{Order}} \bibinfo{volume}{6}, pp.
  \bibinfo{pages}{181--194}, \doi{10.1007/BF02034335}.

\bibitemdeclare{book}{Gib87}
\bibitem{Gib87}
\bibinfo{author}{P.~\surnamestart Gibbins\surnameend} (\bibinfo{year}{1987}):
  \emph{\bibinfo{title}{Particles and Paradoxes: {The} Limits of Quantum
  Logic}}.
\newblock \bibinfo{publisher}{Cambridge UP}, \bibinfo{address}{Cambridge, UK},
\doi{10.1017/CBO9780511570674}.

\bibitemdeclare{book}{Kal83}
\bibitem{Kal83}
\bibinfo{author}{G.~\surnamestart Kalmbach\surnameend} (\bibinfo{year}{1983}):
  \emph{\bibinfo{title}{Orthomodular Lattices}}.
\newblock \bibinfo{publisher}{Academic}, \bibinfo{address}{London}.

\bibitemdeclare{article}{KS67}
\bibitem{KS67}
\bibinfo{author}{S.~\surnamestart Kochen\surnameend} \& \bibinfo{author}{E.~P.
  \surnamestart Specker\surnameend} (\bibinfo{year}{1967}):
  \emph{\bibinfo{title}{The Problem of Hidden Variables in Quantum Mechanics}}.
\newblock {\sl \bibinfo{journal}{J. Math.\ Mech.}} \bibinfo{volume}{17}, pp.
  \bibinfo{pages}{59--87}.

\bibitemdeclare{article}{Mar70}
\bibitem{Mar70}
\bibinfo{author}{E.~L. \surnamestart Marsden\surnameend}
  (\bibinfo{year}{1970}): \emph{\bibinfo{title}{The commutator and solvability
  in a generalized orthomodular lattice}}.
\newblock {\sl \bibinfo{journal}{Pacific J.\ Math}} \bibinfo{volume}{33}, pp.
  \bibinfo{pages}{357--361}, \doi{10.2140/pjm.1970.33.357}.

\bibitemdeclare{book}{vN55}
\bibitem{vN55}
\bibinfo{author}{J.~\surnamestart von Neumann\surnameend}
  (\bibinfo{year}{1955}): \emph{\bibinfo{title}{Mathematical Foundations of
  Quantum Mechanics}}.
\newblock \bibinfo{publisher}{Princeton UP}, \bibinfo{address}{Princeton, NJ}.
\newblock \bibinfo{note}{[English translation of {\it Mathematische Grundlagen
  der Quantenmechanik} (Springer, Berlin, 1932)]}.

\bibitemdeclare{article}{84QC}
\bibitem{84QC}
\bibinfo{author}{M.~\surnamestart Ozawa\surnameend} (\bibinfo{year}{1984}):
  \emph{\bibinfo{title}{Quantum measuring processes of continuous
  observables}}.
\newblock {\sl \bibinfo{journal}{J. Math.\ Phys.}} \bibinfo{volume}{{25}}, pp.
  \bibinfo{pages}{79--87},  \doi{10.1063/1.526000}.

\bibitemdeclare{article}{04URN}
\bibitem{04URN}
\bibinfo{author}{M.~\surnamestart Ozawa\surnameend} (\bibinfo{year}{2004}):
  \emph{\bibinfo{title}{Uncertainty relations for noise and disturbance in
  generalized quantum measurements}}.
\newblock {\sl \bibinfo{journal}{Ann.\ Phys.\ (N.Y.)}} \bibinfo{volume}{311},
  pp. \bibinfo{pages}{350--416}, \doi{10.1016/j.aop.2003.12.012}.

\bibitemdeclare{article}{06QPC}
\bibitem{06QPC}
\bibinfo{author}{M.~\surnamestart Ozawa\surnameend} (\bibinfo{year}{2006}):
  \emph{\bibinfo{title}{Quantum perfect correlations}}.
\newblock {\sl \bibinfo{journal}{Ann.\ Phys.\ (N.Y.)}} \bibinfo{volume}{321},
  pp. \bibinfo{pages}{744--769}, \doi{10.1016/j.aop.2005.08.007}.

\bibitemdeclare{article}{07TPQ}
\bibitem{07TPQ}
\bibinfo{author}{M.~\surnamestart Ozawa\surnameend} (\bibinfo{year}{2007}):
  \emph{\bibinfo{title}{Transfer principle in quantum set theory}}.
\newblock {\sl \bibinfo{journal}{J. Symbolic Logic}} \bibinfo{volume}{72}, pp.
  \bibinfo{pages}{625--648}, \doi{10.2178/jsl/1185803627}.

\bibitemdeclare{article}{11QRM}
\bibitem{11QRM}
\bibinfo{author}{M.~\surnamestart Ozawa\surnameend} (\bibinfo{year}{2011}):
  \emph{\bibinfo{title}{Quantum Reality and Measurement: {A} Quantum Logical
  Approach}}.
\newblock {\sl \bibinfo{journal}{Found. Phys.}} \bibinfo{volume}{41}, pp.
  \bibinfo{pages}{592--607}, \doi{10.1007/s10701-010-9462-y}.

\bibitemdeclare{book}{Red87}
\bibitem{Red87}
\bibinfo{author}{M.~\surnamestart Redhead\surnameend} (\bibinfo{year}{1987}):
  \emph{\bibinfo{title}{Incompleteness, Nonlocality, and Realism: A
  Prolegomenon to the Philosophy of Quantum Mechanics}}.
\newblock \bibinfo{publisher}{Oxford UP}, \bibinfo{address}{Oxford}.

\bibitemdeclare{incollection}{Ta81}
\bibitem{Ta81}
\bibinfo{author}{G.~\surnamestart Takeuti\surnameend} (\bibinfo{year}{1981}):
  \emph{\bibinfo{title}{Quantum set theory}}.
\newblock In \bibinfo{editor}{E.~G. \surnamestart Beltrametti\surnameend} \&
  \bibinfo{editor}{B.~C. \surnamestart van Fraassen\surnameend}, editors: {\sl
  \bibinfo{booktitle}{Current Issues in Quantum Logic}},
  \bibinfo{publisher}{Plenum}, \bibinfo{address}{New York}, pp.
  \bibinfo{pages}{303--322}, \doi{10.1007/978-1-4613-3228-2_19}.

\bibitemdeclare{book}{TZ71}
\bibitem{TZ71}
\bibinfo{author}{G.~\surnamestart Takeuti\surnameend} \& \bibinfo{author}{W.~M.
  \surnamestart Zaring\surnameend} (\bibinfo{year}{1971}):
  \emph{\bibinfo{title}{Introduction to Axiomatic Set Theory}}.
\newblock \bibinfo{publisher}{Springer}, \bibinfo{address}{New York}, 
\doi{10.1007/978-1-4613-8168-6}.

\end{thebibliography}

\end{document}